\documentclass[aps,prb,twocolumn,superscriptaddress,showpacs]{revtex4-1}

\usepackage{graphicx}
\usepackage{calc}
\usepackage{bm}
\usepackage{color}
\usepackage{mathtools}

\begin{document}

\title{Second-harmonic response  in magnetic nodal-line semimetal Fe$_3$GeTe$_2$.}

\author{V.D.~Esin}
\affiliation{Institute of Solid State Physics of the Russian Academy of Sciences, Chernogolovka, Moscow District, 2 Academician Ossipyan str., 142432 Russia}
\author{A.A. Avakyants}
\affiliation{Institute of Solid State Physics of the Russian Academy of Sciences, Chernogolovka, Moscow District, 2 Academician Ossipyan str., 142432 Russia}
\author{A.V.~Timonina}
\affiliation{Institute of Solid State Physics of the Russian Academy of Sciences, Chernogolovka, Moscow District, 2 Academician Ossipyan str., 142432 Russia}
\author{N.N.~Kolesnikov}
\affiliation{Institute of Solid State Physics of the Russian Academy of Sciences, Chernogolovka, Moscow District, 2 Academician Ossipyan str., 142432 Russia}
\author{E.V.~Deviatov}
\affiliation{Institute of Solid State Physics of the Russian Academy of Sciences, Chernogolovka, Moscow District, 2 Academician Ossipyan str., 142432 Russia}

\date{\today}

\begin{abstract}
We experimentally investigate  second-harmonic transverse voltage response to ac electrical current for a magnetic nodal-line semimetal Fe$_3$GeTe$_2$. For zero magnetic field,  the observed  second-harmonic voltage  depends as a square of the   longitudinal current, as it should be expected for non-linear Hall effect.  The magnetic field behavior is found to be  sophisticated: while the first-harmonic response shows the known anomalous Hall hysteresis in FGT,  the second-harmonic Hall voltage is characterized by the pronounced high-field hysteresis  and flat ($B$-independent) region  with curves touching at low fields. The high-field hysteresis strongly depends on the magnetic field sweep rate, so it reflects some slow relaxation process.   For the lowest rates, it is also accomplished by multiple crossing points. Similar shape of the second-harmonic hysteresis is known  for skyrmion spin textures in non-linear optics. Since skyrmions have been demonstrated for FGT by direct visualization techniques,  we can connect the observed high-field relaxation with deformation of the skyrmion lattice. Thus, the  second-harmonic Hall voltage  response    can be regarded as a tool to detect spin textures in transport experiments.  
\end{abstract}

\maketitle

\section{Introduction}

Physics of topological semimetals is a new and growing field of modern condensed matter research~\cite{armitage}. Dirac semimetals are characterized by gapless spectrum, because of band touching in some distinct points, which are the special points of Brillouin zone. In Weyl semimetals every touching point splits  into two Weyl nodes with opposite chiralities due to the time reversal or inversion symmetries breaking. Alternatively, if the band touchings occur along some lines in the three-dimensional Brillouin zone, the material is known as a topological nodal-line semimetal~\cite{armitage,fang,bernevig,zhang}. Topologically protected Fermi arc surface states  are  connecting projections of these nodes on the surface Brillouin zone, which produces  complex spin textures~\cite{jiang15,rhodes15,wang16} on the surface due to the spin-momentum locking~\cite{Sp-m-lock}.

One of the promising candidates for magnetic nodal-line semimetal~\cite{kim} is a van der Waals ferromagnet Fe$_3$GeTe$_2$ (FGT)~\cite{2dFM1,2dFM2,2dFM3,anisotrop,kerr,fgttc}. Experimentally, FGT shows large anomalous Hall~\cite{kim,ahe1} and Nernst~\cite{Nernst} effects, topological Hall effect~\cite{PTHE}, giant tunneling magnetoresistance~\cite{magres} and Kondo lattice physics~\cite{kondo}. Also, the nontrivial topological spin textures - magnetic skyrmions - have been   demonstrated~\cite{skyrmion,skyrmion2} in FGT, in addition to the conventional labyrant domain structure~\cite{fgtdomain1,fgtdomain2}.

FGT magnetization can be investigated by different techniques, but not all of them are  sensitive to the relatively small number of spins at the surface of topological semimetals. For example, a typical anomalous Hall hysteresis loop mostly reflects the bulk magnetization behavior~\cite{CoSnS_spin glass}. On the other hand,  the harmonic Hall analysis~\cite{hayashi,vlietstra,avci,mac,chen,schippers,bauer} is a known transport  technique to study  spin textures in different materials.  In general, it  is a part of a broad approach, which is known also  in nonlinear optics~\cite{shg,hall,hall2,skyrmion0}, where this technique was demonstrated for optical investigations of skyrmion structures~\cite{skyrmion0}. 
 
An important example of the  harmonic Hall response in topological materials is the non-linear Hall (NLH) effect~\cite{sodemann}, which is predicted as a transverse current at both zero and twice the    frequency~\cite{deyo,golub,moore,low,nlhe2,nlhe3,nlhe4,nlhe5,nlhe6,nlhe7,nlhe8,nlhe9,nlhe10,nlhe11,nlhe12,nlhe13}.  NLH effect has been experimentally demonstrated for monolayer transitional metal dichalcogenides~\cite{ma,kang} and for three-dimensional Weyl and Dirac semimetals~\cite{esin,c_axis} as a second-harmonic Hall voltage in zero magnetic field. For FGT, one can also expect that the harmonic Hall analysis  in finite magnetic fields can be a powerful tool to investigate spin textures.

 Here, we experimentally investigate  second-harmonic transverse voltage response to ac electrical current for a magnetic nodal-line semimetal Fe$_3$GeTe$_2$. For zero magnetic field,  the observed  second-harmonic transverse voltage  depends as a square of the  longitudinal current, as it should be expected for non-linear Hall effect. The magnetic field behavior is found to be  sophisticated for the magnetic topological semimetal FGT, where spin textures have a significant effect on the  second-harmonic Hall voltage.

\section{Samples and technique}

\begin{figure}
\includegraphics[width=1\columnwidth]{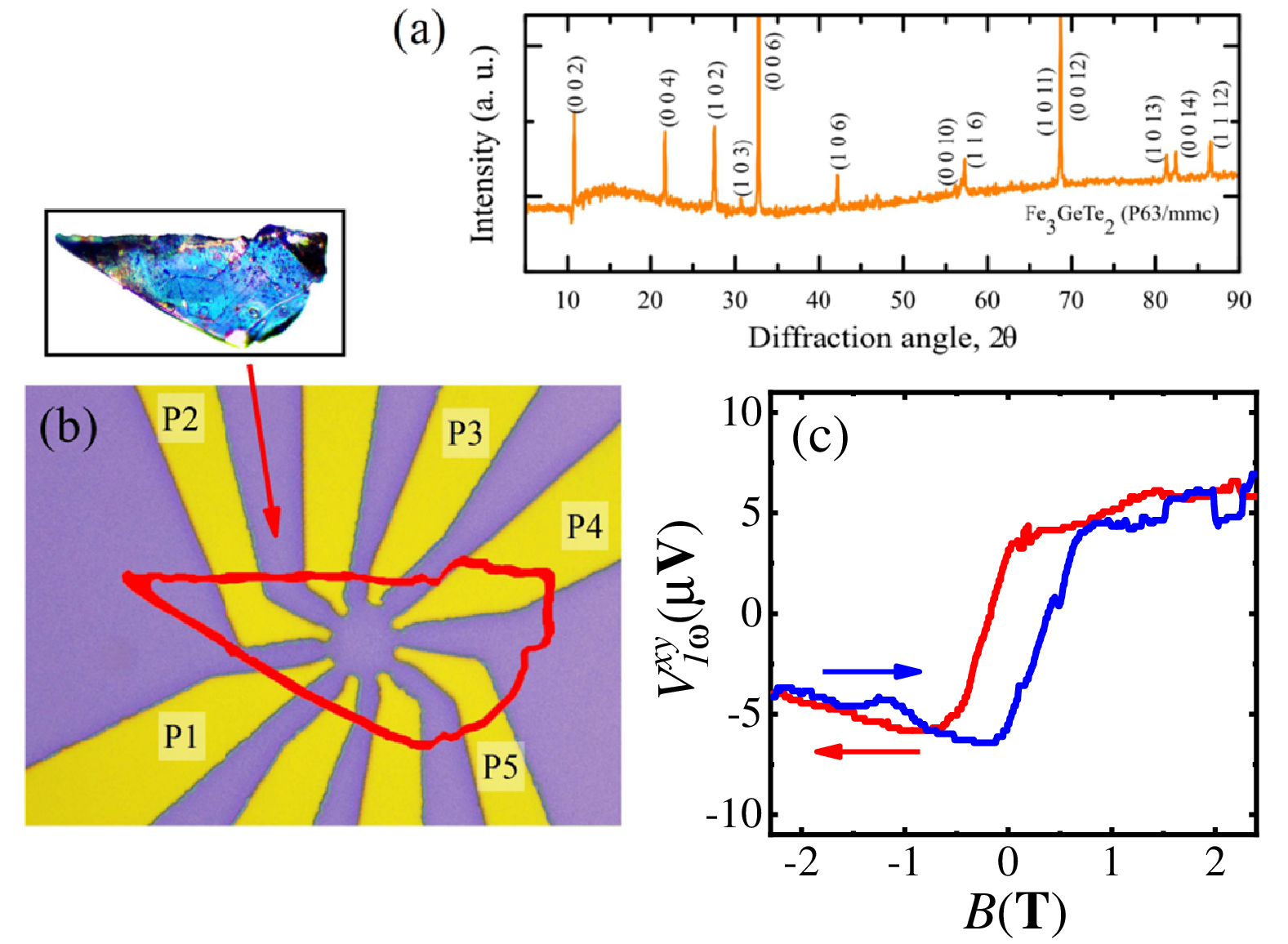}
\caption{(Color online) (a) X-ray diffraction pattern (Cu K$\alpha$1 radiation, $\lambda$ = 1.540598 \AA), which confirms single-phase Fe$_3$GeTe$_2$ with P63/mmc (194) space group (a = b = 3.991(1) \AA, c = 16.33(3) \AA).
(b) Optical image of the Au leads  on the insulating SiO$_2$ substrate. 100 nm thick, 5~$\mu\mbox{m}$ separated  leads form a circle in the central part with 18~$\mu$m diameter.  A small (about 100~$\mu\mbox{m}$ size and 0.5~$\mu\mbox{m}$ thick) single-crystal FGT ﬂake is transferred to the leads, as it is depicted by the arrow. The ac current is applied between P1 and P4 contacts, while the transverse (Hall) voltage $V^{xy}$ is measured between the  P3 and P5 potential probes.  (c) A large anomalous Hall effect as a first-harmonic  $V_{1\omega}^{xy}$ hysteresis loop in normal magnetic field, which confirms the known magnetic properties of FGT~\cite{PTHE,infgt}. The  arrows indicate the magnetic field sweep directions. }
\label{sample}
\end{figure}

Fe$_3$GeTe$_2$ was synthesized from elements in evacuated silica ampule in a two-step process. At the first step, the load was heated up to 470$^{\circ}$ C at 10 deg/h rate and the ampule was held at this temperature for 50 h. At the second step, the temperature was increased up to 970$^{\circ}$ C with the same rate. After 140 h exposure, the ampule was cooled down to the room temperature at 5 deg/h rate. X-ray diffraction data indicates, that the iron tellurides FeTe and FeTe$_2$ were also found in the  material, in addition to the expected Fe$_3$GeTe$_2$ compound.

To obtain Fe$_3$GeTe$_2$ single crystals, the synthesized mixture was sealed in evacuated silica ampule with some admixture of iodine. The transport reaction was carried out for 240 h with temperatures 530$^{\circ}$ C and 410$^{\circ}$ C in hot and cold zones, respectively. Afterward, the ampule was quenched in a liquid nitrogen. Water-solvable iron and tellurium iodides were removed in hot distilled water from the obtained Fe$_3$GeTe$_2$ single crystals, so the X-ray diffraction analysis confirms single-phase Fe$_3$GeTe$_2$ with P63/mmc (194) space group (a = b = 3.991(1) \AA, c = 16.33(3) \AA), see Fig.~\ref{sample}(a).  The known structure model~\cite{x-ray-FGT} Fe$_3$GeTe$_2$ is refined with single crystal X-ray diffraction measurements (Oxford diffraction Gemini-A, MoK$\alpha$). The Fe$_3$GeTe$_2$ composition is also verified by energy-dispersive X-ray spectroscopy.

Despite FGT is ferromagnetic even for two-dimensional monolayer samples, topological semimetals are essentially three-dimensional objects~\cite{armitage}. Thus, we have to select  relatively thick (above 0.5~$\mu$m) FGT single crystal flakes, which also ensures sample homogeneity for correct determination of xx- and xy- voltage responses. Thick flakes requires special contact preparation technique: the mechanically exfoliated flake is transferred on the Au leads pattern, which is defined on the standard oxidized silicon substrate by lift-off technique, as depicted in Fig.~\ref{sample} (b). The transferred flake is shortly pressed  to the leads by another oxidized silicon substrate, the latter is removed afterward. This procedure provides transparent FGT-Au junctions (below 1~Ohm resistance), stable in different cooling cycles, which has been verified  before for a wide range of materials~\cite{cdas,cosns,black,timnal,infgt}. As an additional advantage, the relevant FGT surface with Au contacts (the bottom one) is protected from any contamination by SiO$_2$ substrate. 

We investigate  transverse (xy-)  first- and second-harmonic  voltage responses by standard four-point lock-in technique.  The ac current is applied between P1 and P4 contacts in Fig.~\ref{sample} (b), while the transverse (Hall) voltage $V^{xy}$   is measured between the  P3 and P5 potential probes. Also, the longitudinal $V^{xx}$ component can be measured between the P2 and P3. The Curie temperature of bulk FGT crystals~\cite{PTHE} is  about $\approx 220$~K, so the measurements are performed at the liquid helium temperatures. Similar results are obtained for several samples in different cooling cycles.

\begin{figure}
\includegraphics[width=\columnwidth]{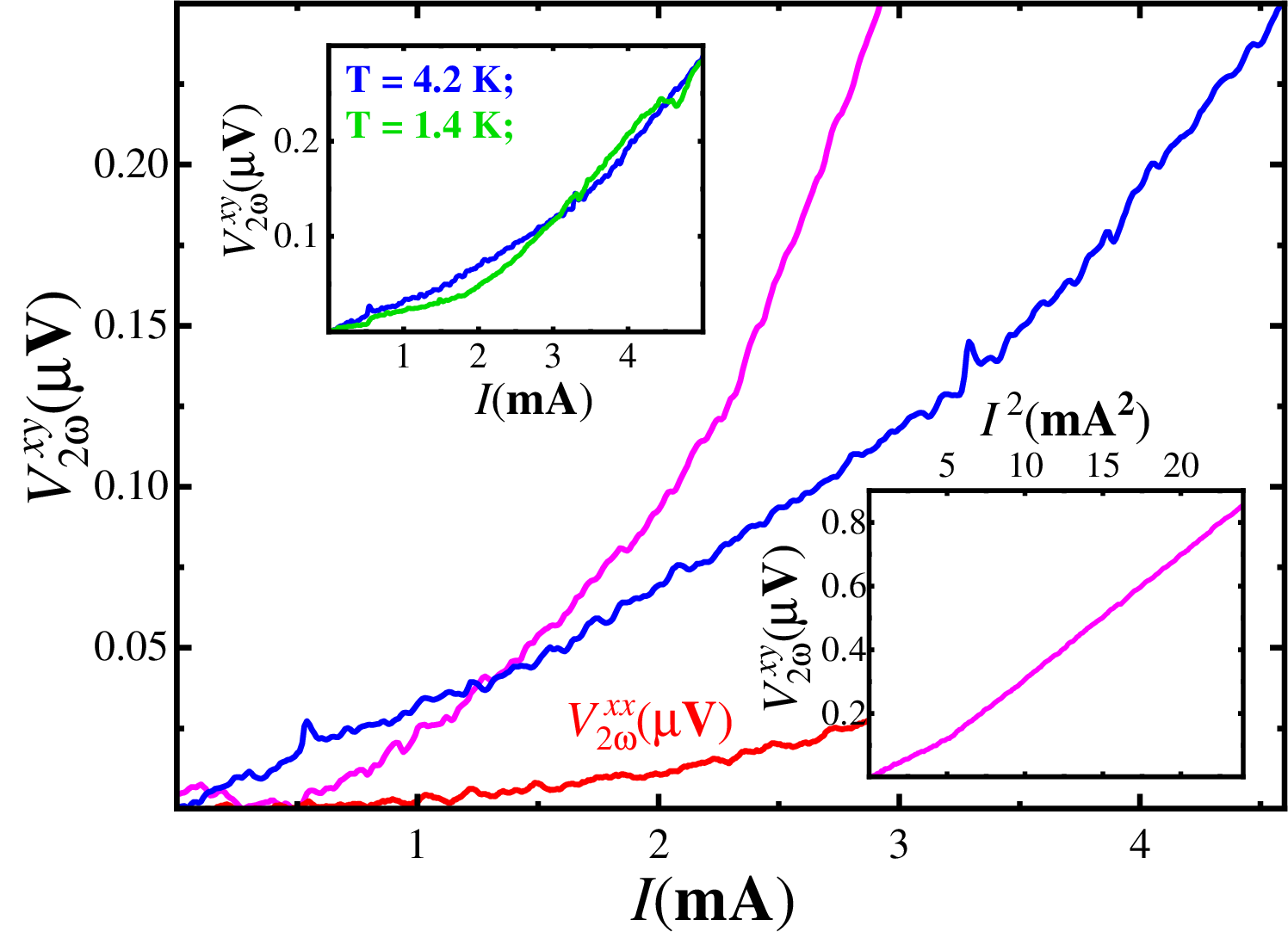}
\caption{(Color online) Zero magnetic field. Typical behavior of the transverse  second-harmonic voltage component $V_{2\omega}^{xy}\sim I^2$, as it should be expected for the non-linear Hall effect~\cite{ma,kang,esin,c_axis}. The longitudinal  second-harmonic voltage $V_{2\omega}^{xx}$  is one order of magnitude smaller. The data are presented for two different samples (blue and magenta curves, respectively)  at 4.2~K.  For clarity, $V_{2\omega}^{xx}$ (red) is only shown for the sample with the highest $V_{2\omega}^{xy}$ values (the magenta curve), the bottom inset demonstrates the square-law $\sim I^2$ dependence for this sample. The upper inset shows $V_{2\omega}^{xy}$ curves for two different temperatures 4.2 and 1.4~K, there is practically no difference in this temperature range.}
\label{VAC}
\end{figure}

\section{Experimental results}

We confirm the correctness of the experimental conditions by demonstrating  first-harmonic  $V_{1\omega}^{xy}$ anomalous Hall hysteresis loop, see  Fig.~\ref{sample} (c).   A large anomalous Hall effect  manifests itself as non-zero Hall voltage in zero magnetic field, which is determined by the bulk magnetization direction.  The observed first-harmonic  hysteresis loop  well corresponds to the known anomalous Hall effect in FGT~\cite{PTHE,infgt}. 
  
In zero external magnetic field, Fig.~\ref{VAC} shows typical behavior of the non-linear Hall effect~\cite{ma,kang,esin,c_axis}  as a quadratic transverse Hall-like response $V_{2\omega}^{xy}$ to ac excitation current $I$ for two different samples. The $\sim I^2$ dependence  is directly demonstrated in the bottom inset to Fig.~\ref{VAC}. The longitudinal  second-harmonic voltage $V_{2\omega}^{xx}$ is one order of magnitude smaller, which confirms well-defined Au leads geometry and a homogeneous FGT flake. There is no noticeable temperature dependence in the 1.4-4.2~K range, see the upper inset in  Fig.~\ref{VAC}, since the FGT spectrum  is well-established much below  the $\approx 220$~K Curie temperature~\cite{PTHE}. Thus, we observe non-linear Hall effect for magnetic nodal-line semimetal FGT. 

\begin{figure}
\includegraphics[width=\columnwidth]{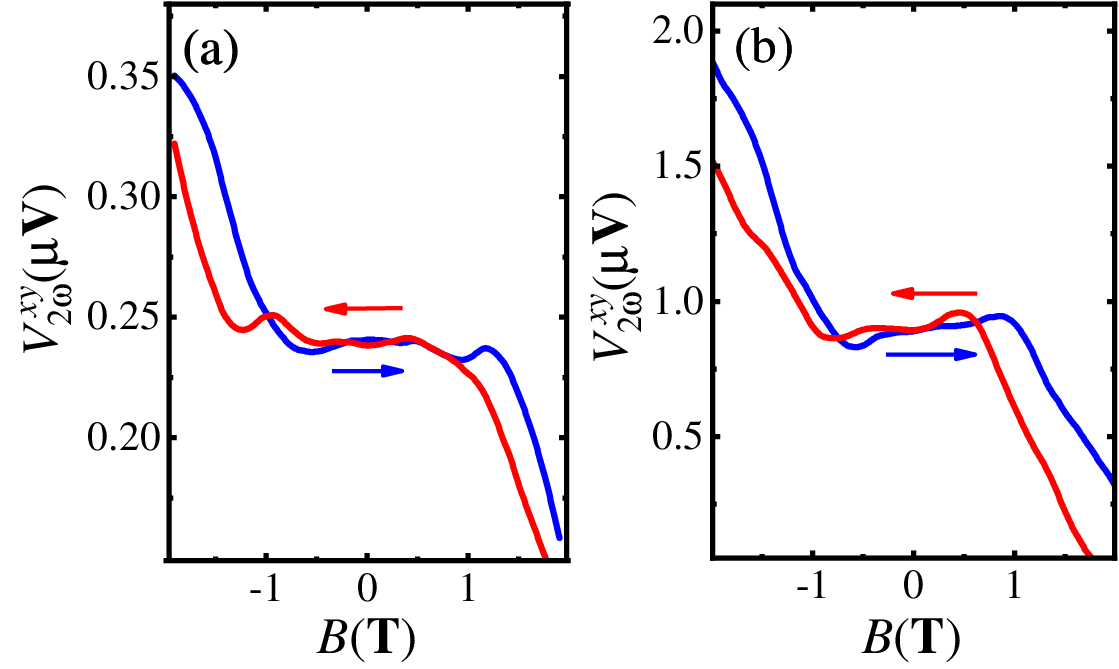}
\caption{(Color online) Asymmetric $V_{2\omega}^{xy}(B)$ magnetic field dependence, which is qualitatively similar for two different samples with strongly different $V_{2\omega}^{xy}$ values ((a) and (b), respectively, for the fixed ac current $I=4.5$~mA ). The expected linear $\sim B$ field contribution  
is accomplished by pronounced high-field hysteresis and flat ($B$-independent) region  with curves touching at low fields, within $\pm 1$~T. The hysteresis is defined by magnetic field sweep direction, as indicated by arrows of the same color. The curves are  shown  for high, 8~mT/s sweep rate for normal orientation of the magnetic field, at 4.2~K temperature.}
\label{MagnetPerpendicullar}
\end{figure}

In the magnetic field, Fig.~\ref{MagnetPerpendicullar} shows asymmetric $V_{2\omega}^{xy}(B)$ dependences for two opposite field sweep directions. The linear $\sim B$ field contribution can be expected~\cite{mandal,zyuzin} for the NLH effect signal, which is an origin of $V_{2\omega}^{xy}(B)$ asymmetry, as it was experimentally confirmed for non-magnetic topological semimetals~\cite{esin}. However, Fig.~\ref{MagnetPerpendicullar} shows much more sophisticated behavior for  ferromagnetic FGT flakes. We indeed observe linear field dependence in high magnetic fields, which is accomplished by pronounced hysteresis and flat ($B$-independent) region  with curves touching at low fields, within $\pm 1$~T. This  $V_{2\omega}^{xy}(B)$ behavior is qualitatively similar for two different samples with strongly different $V_{2\omega}^{xy}$ values in Fig.~\ref{MagnetPerpendicullar} (a) and (b).

The hysteresis amplitude  depends on the magnetic field sweep rate, while the hysteresis itself is present even for the lowest rates, see Fig.~\ref{MagnetParallel}. We observe pronounced hysteresis in the $V_{2\omega}^{xy}(B)$ curves with the magnetic field sweep direction  for high, 8~mT/s sweep rate in Fig.~\ref{MagnetPerpendicullar}, while it is much smaller for 1~mT/s in Fig.~\ref{MagnetParallel}, so the high-field hysteresis reflects some slow relaxation process.  At the lowest sweep rates, multiple crossing points are also observed, so  the details of the hysteresis loop differ for two samples in Fig.~\ref{MagnetParallel} (a) and (b). Multiple crossing points usually reflect inhomogeneous magnetization process for spin textures in the sample~\cite{hall, hall2,skyrmion0}. Inset to Fig.~\ref{MagnetParallel} shows non-linear planar Hall effect~\cite{NLHE_planar} as the asymmetric $V_{2\omega}^{xy}(B)$ behavior for the in-plane magnetic field orientation. Despite the results are qualitatively similar for  two field orientations, the hysteresis is more pronounced in the parallel field even for the lowest (1~mT/s) sweep rate, while the magnetic field dependence itself is weaker for the parallel field.

\section{Discussion} \label{disc}

\begin{figure}
\includegraphics[width=\columnwidth]{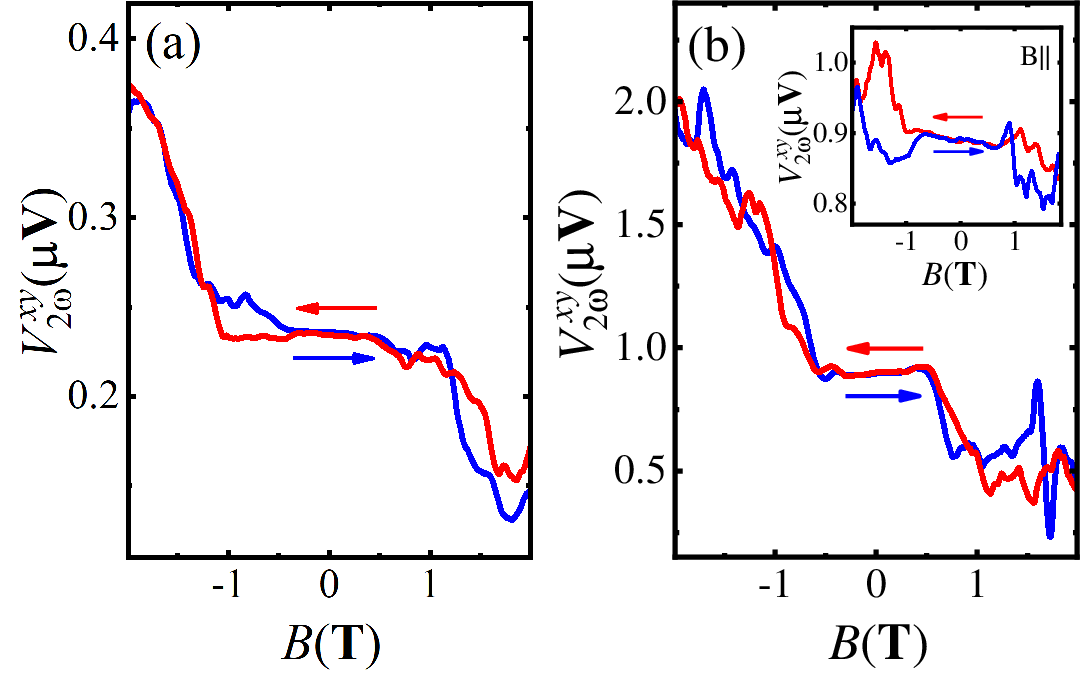}
\caption{ (Color online)  $V_{2\omega}^{xy}(B)$ curves for the lowest, for 1~mT/s sweep rate, (a) and (b) panels are  for two  samples from Fig.~\ref{MagnetPerpendicullar}, respectively. The hysteresis amplitude  is significantly smaller in this case, there are multiple crossing points for the curves, which usually reflects some inhomogeneous magnetization process~\cite{hall, hall2,skyrmion0}, so the details of the hysteresis loops differ for two samples. The curves are obtained at 4.2~K for normal orientation of the magnetic field and the fixed ac current $I=4.5$~mA. Inset  shows 
much more pronounced hysteresis for the in-plane magnetic field orientation for the same (1~mT/s) sweep rate, while the magnetic field dependence itself is weaker in this case. Arrows indicate the magnetic field sweep direction. }
\label{MagnetParallel}
\end{figure}

As a result, while the NLH effect could be expected for   Fe$_3$GeTe$_2$ in zero magnetic field, the sophisticated behavior of $V_{2\omega}^{xy}(B)$ requires consistent explanation.

In principle, second-harmonic hysteresis could also arise from Joule heating $\sim R I^2$ of the sample with low thermoconductance~\cite{avci}. On the other hand,  sample magnetoresistance $R(B)$ is not sensitive to the magnetic field sign, so any thermoelectric effects should be symmetric in magnetic field~\cite{avci,esin,thermo2w,wees}, in contrast to experimental asymmetric (odd) magnetic field dependences in Fig.~\ref{MagnetPerpendicullar}. The hysteresis can not also originate from the experimental equipment, since we never observed  $V_{2\omega}^{xy}(B)$  hysteresis for non-magnetic samples in similar experiments~\cite{esin,thermo2w}. 

The hysteresis in Hall voltage $V_{1\omega}^{xy}$ is known for  FGT flakes in the external magnetic field~\cite{PTHE,infgt} as   anomalous Hall effect~\cite{armitage}. For topological materials, it is usually regarded as the indication of a magnetic topological phase, as supported, e.g.,  by the topological-insulator-multilayer model, so the one-dimensional Chern edge states form the two-dimensional surface states~\cite{armitage}. Irrespective of the particular mechanism, anomalous Hall effect reflects the bulk magnetization of the ferromagnetic FGT flakes~\cite{PTHE,infgt}: $V_{1\omega}^{xy}$ Hall voltage changes its sign if the bulk magnetization is reversed by the external magnetic field. Experimentally, the Hall voltage hysteresis well corresponds to the $M(H)$ magnetization reversal curves, see, e.g.,  Ref.~\cite{CoSnS_spin glass}.

These considerations can not be directly applied to the  second-harmonic Hall voltage component, since the non-linear Hall effect arises from the Berry curvature in momentum space~\cite{sodemann}. In the simplified picture, an a.c. excitation current generates the effective sample magnetization, which leads to the Hall effect  in zero external magnetic field. Hall voltage is therefore  proportional to the square of the excitation current, so it can be detected as the second-harmonic transverse voltage component $V_{2\omega}^{xy}$, as we observe in Fig.~\ref{VAC}. Another possible contribution to the non-linear Hall effect is skew scattering with nonmagnetic impurities in time-reversal-invariant noncentrosymmetric materials~\cite{skew}, but it hardly be applied to the ferromagnetic FGT semimetal.

Theoretically predicted $V_{2\omega}^{xy}$ sensitivity to the external magnetic field~\cite{mandal,zyuzin} indicates, that $V_{2\omega}^{xy}$ should also be  sensitive to the internal bulk FGT magnetization. From the comparison of Figs.~\ref{sample} (c) and~\ref{MagnetPerpendicullar}, the internal magnetization is dominant within $\pm 1$~T, the region of the $V_{1\omega}^{xy}$ hysteresis loop, while the linear $V_{2\omega}^{xy}(B)\sim B$ dependence on the external field appears beyond this region in Fig.~\ref{MagnetPerpendicullar}. This well describes the flat region in the experimental $V_{2\omega}^{xy}(B)$ curves, but the high-field hysteresis seems to have a different origin, because the internal magnetization is not sensitive to the magnetic field outside $\pm 1$~T, see Fig.~\ref{sample} (c). On the other hand, similar hysteresis is known for nonlinear optics, where a second-harmonic signal is a powerful method to analyze skyrmion spin textures in magnetic materials~\cite{hall, hall2,skyrmion0}.  
 In particular, multiple crossing points  reflect inhomogeneous magnetization, which is  a fingerprint of spin textures~\cite{hall, hall2,skyrmion0}. 

For the FGT semimetal, a high-density lattice of hexagonally packed  skyrmions can be induced by a simple cooling process~\cite{skyrmion2,skyrmion}.   By Bitter decoration technique~\cite{vinnikov1, vinnikov2}, we also confirmed the  labyrant domain structure~\cite{fgtdomain1,fgtdomain2}  for our FGT  samples, as well as the  hexagonally packed  skyrmions~\cite{skyrmion2}.  Also, the  characteristic bow-tie magnetic hysteresis loops are shown for our FGT samples~\cite{cosnsmag}, which are  usually ascribed  to the skyrmions. Deformation of the skyrmion lattice can be responsible for the observed hysteresis with the magnetic field sweep direction. Indeed, skyrmions appear in magnetic materials due to the  Dzyaloshinsky-Moriya interaction~\cite{moree}.  The competition between the perpendicular magnetic anisotropy and magnetic dipole-dipole interaction is crucial for skyrmions, so the spin textures should be sensitive to the external magnetic field.  

This conclusion is strongly confirmed by the $V_{2\omega}^{xy}(B)$ dependence  on the field sweep rate in Figs.~\ref{MagnetPerpendicullar} and~\ref{MagnetParallel}.  At the lowest rate, multiple crossing points reflect inhomogeneous magnetization  in the presence of skyrmion structures in a good correspondence with the known behavior  in optics~\cite{hall, hall2,skyrmion0}. The particular skirmion distribution is obviously different for different samples, so the details of the  shape of the hysteresis loop differs in Fig.~\ref{MagnetParallel} (a) and (b). On the other hand, the curves are very similar for high sweep rates, where the hysteresis only reflects the deformation of the skyrmion lattice. 

For the in-plane magnetic field, one can also expect $B$-like correction to the $V^{xy}_{2\omega}(B)$ dependence~\cite{zyuzin,NLHE_chiral}. The surface spin textures should be sensitive to the direction of the external magnetic field, so  the slow relaxation is more pronounced in the inset to  Fig.~\ref{MagnetParallel}.  The central flat region is obviously wider (approximately from -1~T to 1~T) for the parallel field orientation in the inset to Fig.~\ref{MagnetParallel} (b), in contrast to the  -0.5~T to 0.5~T region for normal field. This difference is consistent with the known magnetic anisotropy in FGT, see the data in Ref.~\cite{infgt} for our samples.

 This confirms our interpretation, so the second-harmonic Hall response $V^{xy}_{2\omega}(B)$   is  a powerful tool to detect spin textures in magnetic nodal-line semimetals because of the sensitivity of the second-harmonic response (non-linear Hall signal) to the inhomogeneous magnetization processes,  in contrast to the first-harmonic (conventional anomalous Hall one).

\section{Conclusion}

As a conclusion, we observe sophisticated magnetic field behavior of the second-harmonic Hall voltage response: while the first-harmonic signal shows the known anomalous Hall hysteresis in FGT,  the second-harmonic Hall voltage is characterized by the pronounced high-field hysteresis  and flat ($B$-independent) region in $V^{xy}_{2\omega}(B)$  with curves touching at low fields. The high-field hysteresis reflects some slow relaxation process, so it strongly depends on the magnetic field sweep rate. For the lowest rates, it is also accomplished by multiple crossing points. The low-field curves touching and the shape of the second-harmonic hysteresis with multiple crossing points are known  for skyrmion spin textures in non-linear optics. Since skyrmions have been demonstrated for FGT by direct visualization techniques,  we can connect the observed high-field relaxation with deformation of the skyrmion lattice. This conclusion is confirmed by the $V^{xy}_{2\omega}(B)$ sensitivity to the direction of the external magnetic field, as it should be expected for surface spin textures. Thus, the second-harmonic Hall response hysteresis   can be regarded as the  manifestation of Fe$_3$GeTe$_2$ skyrmion structures in transport experiments.

\acknowledgments

We wish to thank S.S~Khasanov for X-ray sample characterization.  We gratefully acknowledge financial support  by the  RF State task.

\end{document}